# Unusual effects of manual grinding and subsequent annealing process observed in Gd$_{5.09}$Ge$_{2.03}$Si$_{1.88}$ compound

A. M. G. Carvalho [1,*], C. S. Alves [2], P. V. Trevizoli [2], A. O. dos Santos [3],

S. Gama [4], A. A. Coelho [5]

[1] *Laboratório Nacional de Luz Síncrotron (LNLS), Centro Nacional de Pesquisa em Energia e Materiais (CNPEM), CEP 13083-100, Campinas, SP, Brazil*

[2] *Departamento de Engenharia Mecânica, UEM, 87020-900, Maringá, PR, Brazil.*

[3] *Centro de Ciências Sociais, Saúde e Tecnologia, UFMA, Imperatriz, MA, Brazil.*

[4] *Departamento de Ciências Exatas e da Terra, UNIFESP, Diadema, SP, Brazil.*

[5] *Instituto de Física Gleb Wataghin, UNICAMP, 13083-859, Campinas, SP, Brazil.*

*\* Corresponding author. E-mail: alexandre.carvalho@lnls.br*

## Abstract

The Gd$_{5.09}$Ge$_{2.03}$Si$_{1.88}$ compound, as well as other magnetocaloric materials, certainly will not be used in their un-manufactured as-cast condition in future magnetic refrigeration applications or other devices. In this work, we have studied the Gd$_{5.09}$Ge$_{2.03}$Si$_{1.88}$ compound processed in different ways, mainly, the as-cast powder, the annealed powder and the pressed and sintered powder. The annealed powder (1370 K / 20 hours) does not present the monoclinic phase and the first-order-magneto-structural transition observed in the as-cast powder. The pressed and sintered powder also do not present the first-order transition. Furthermore, the compacting pressure shifts the second-order-magnetic transition to lower temperatures. The behavior of cell parameters as a function of the compacting pressure indicates that T$_C$ is directly affected by parameter *c* change.

## 1. Introduction

In 1997, V. K. Pecharsky and K. A. Gschneidner Jr. reported the giant magnetocaloric effect in the $Gd_5Ge_2Si_2$ compound [1] and in other compounds of the family $Gd_5(Ge_xSi_{1-x})_4$ [2]. As shown in Ref. 2, from this family of materials, the $Gd_5Ge_2Si_2$ compound is not the one that presents the greatest magnetocaloric effect (MCE) described by the quantity $\Delta S_T$ (isothermal variation of entropy). However, it presents this effect around room temperature. This fact makes the compound $Gd_5Ge_2Si_2$ a potential candidate to be used as a refrigerant material in a magnetic refrigerator that eventually will substitute the conventional refrigerators and air-conditioning devices.

According to A. O. Pecharsky *et al.* [3], the as-cast $Gd_5Ge_2Si_2$ compound presents two magnetic transitions, one of first-order nature around 277 K, due to a majority phase, and another of second-order nature around 300 K, due to a minority phase. Repeating the preparation procedures from Ref. 3, but using commercial (low purity) gadolinium, we do not obtain the same results. Our as-prepared $Gd_5Ge_2Si_2$ compound presents only the second-order transition at 300 K, such as has been reported for other commercial-purity-Gd-as-cast samples of $Gd_5Ge_2Si_2$ [4,5]. To obtain the first-order transition, the one of lowest temperature, it is necessary to carry out heat treatments [4].

When some of us performed metallographic and WDS (Wavelenght Dispersive Spectroscopy) analyses, we have verified that our $Gd_5Ge_2Si_2$ compound presented two crystallographic phases: the majority phase, with stoichiometry $Gd_{5.09}Ge_{2.03}Si_{1.88}$, and the minority one, with stoichiometry $Gd_{4.59}Ge_{1.74}Si_{2.67}$ [6]. Then, we have decided to prepare a compound with the stoichiometry of the majority phase of $Gd_5Ge_2Si_2$, i. e., $Gd_{5.09}Ge_{2.03}Si_{1.88}$. As happens with the $Gd_5Ge_2Si_2$ stoichiometry, this new compound is not a single phase sample. Nevertheless, the most important experimental observation is that, even in the as-cast condition, the $Gd_{5.09}Ge_{2.03}Si_{1.88}$ compound presents the two magnetic transitions, first and second-order ones, as it occurs in $Gd_5Ge_2Si_2$ compound in Ref. [3]. More important, the first-order transition is much stronger than the second-order one.

It is important to remark that the $Gd_{5.09}Ge_{2.03}Si_{1.88}$ compound, as well as other magnetocaloric materials, certainly will not be used in their un-manufactured as-cast condition in future magnetic refrigeration applications or other devices. To use these materials as magnetic active regenerators, it will be necessary to process them into forms dictated by the design of the machine, as wires, thin plates, spheres or other convenient shape. Since this particular compound is very hard and brittle, we considered the use of the techniques of powder metallurgy in order to prepare simple shapes of the material. Our first choice was to grind the as-cast material, sieve it into convenient powder particle sizes, press a small disc of the material (what we will refer to as a tablet) and then, sinter it in order to have the final tablet into a consolidated piece.

It is worth mentioning that $Gd_{5.09}Ge_{2.03}Si_{1.88}$ compound has been studied by some of us using different techniques, such as: magnetization measurements using commercial magnetometers [7]; electron spin resonance [7,8]; magnetoacoustic technique [9,10]. Besides, it has been studied a few magnetic properties of sintered tablets of $Gd_{5.09}Ge_{2.03}Si_{1.88}$ compound, prepared from powder with different granulometries [11]. In the present work, we present the study of magnetic and structural properties of $Gd_{5.09}Ge_{2.03}Si_{1.88}$ compound processed in different ways, mainly, the as-cast powder, the annealed powder and the pressed and sintered powder.

## 2. Experimental details

The $Gd_{5.09}Ge_{2.03}Si_{1.88}$ compound (nominal composition) was prepared using Gd, 99.9 wt.% pure, Si and Ge, 99.99+ wt.% pure, by arc-melting for three times to guarantee the sample homogeneity. The mass of the sample was 5 g. Part of the sample was then manually ground using ceramic mortar and pestle. The resulting powder was sieved to obtain powder particles smaller than 50 μm. Part of this powder was pressed in tablet form using a hardened tool steel die. The compacting pressures were applied using a conventional hydraulic press and measured using a calibrated strain gauge force cell, able to accurately measure forces up to 50 kN. We used different compacting pressures (130, 173, 217 and 260 MPa) to obtain the tablets. This range of compacting pressures is typical of the pressures used to obtain green bodies with enough mechanical strength that allow their easy manipulation. The original powder and the tablets were encapsulated into quartz ampoules under Ar inert atmosphere and sintered at 1370 K for 20 hours using a conventional resistive tubular furnace.

Magnetic measurements were carried out using a commercial magnetometer (Quantum Design, model MPMS XL). Magnetization versus temperature data were obtained always in a zero-field cooling – field warming process, i. e., the sample is cooled from room temperature with no applied magnetic field and, when in the ferromagnetic regime, the magnetic field is applied. Then, we perform the magnetization measurements increasing and then decreasing the temperature.

X-ray data for these samples were collected using $CuK_\alpha$ radiation with a graphite diffracted beam monochromator, at room temperature (around 300 K), using a Philips diffractometer (PW 1710). The crystalline structure refinement of each observed phase in the samples was obtained through the Rietveld method, using GSAS.

## 3. Results and discussions

The as-cast $Gd_{5.09}Ge_{2.03}Si_{1.88}$ compound in bulk presents two magnetic transitions (FIG. 1). Around 275 K, there is a first-order-magneto-structural transition from a ferromagnetic orthorhombic phase to a paramagnetic monoclinic phase. There is also a 4 K thermal hysteresis typical of that kind of magnetocaloric materials. Around 302 K, the compound presents a second-order-magnetic transition due to a second stoichiometric phase. This transition is also from a ferromagnetic to a paramagnetic phase. It is worth noting that, grinding the sample manually into powder form (50 µm or less), the magnetization profile around the transition is not much changed, but enough to be noticed, as seen in FIG. 1 (the curves are normalized). However, the first-order-magneto-structural transition is shifted about −3 K. It strongly indicates that a permanent lattice distortion was induced in the powder particles. Besides, it seems that this grinding process subtly favors the first-order transition, depleting the second-order transition. The tiny magnetic transition observed in the as-cast powder around 300 K is attributed to the orthorhombic $Gd_5(Ge,Si)_4$ phase [12], which is not observed with x-ray diffraction (Table 1). However, annealing the powder at 1370 K for 20 hours, the first-order transition disappears, remaining a second-order transition around 298 K (FIG. 1). It is very important to mention that these annealing conditions do not destroy the first-order transition in bulk samples of $Gd_{5.09}Ge_{2.03}Si_{1.88}$.

The process of pressing the as-cast powder in tablet form, using different compacting pressures, and sintering the tablets at 1370 K for 20 hours also leads to the destruction of the first-order transition (FIG. 2). Besides, increasing the compacting pressure, the second-order transition shifts to lower temperatures. Although the tablets were pressed in air, the oxygen contamination is irrelevant and could not induce these behaviors. According to W. Wu *et al*. [13], the second-order transition temperature of a similar compound is not reduced with the oxygen content and a large amount of oxygen is necessary to destroy the first-order transition.

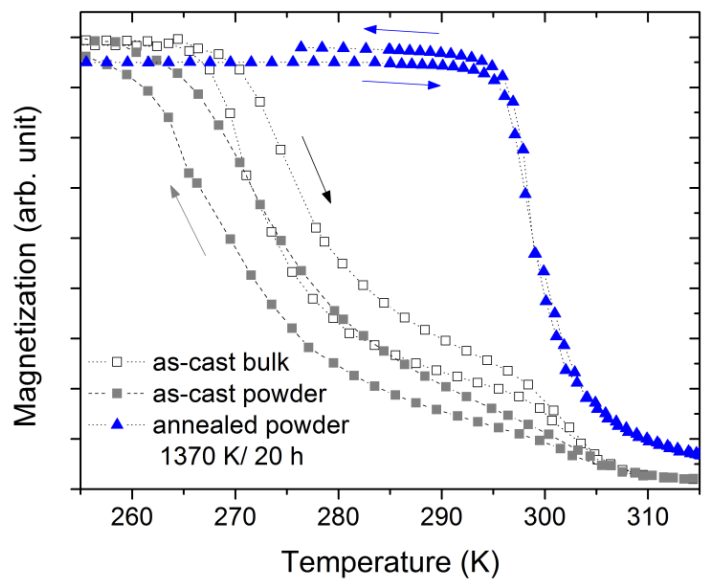

FIG. 1. Magnetization profiles of the as-cast bulk (open squares), the as-cast powder (solid squares) and the annealed powder (solid triagles) of $Gd_{5.09}Ge_{2.03}Si_{1.88}$ compound obtained at 0.02 T. The curves were measured increasing and decreasing temperature.

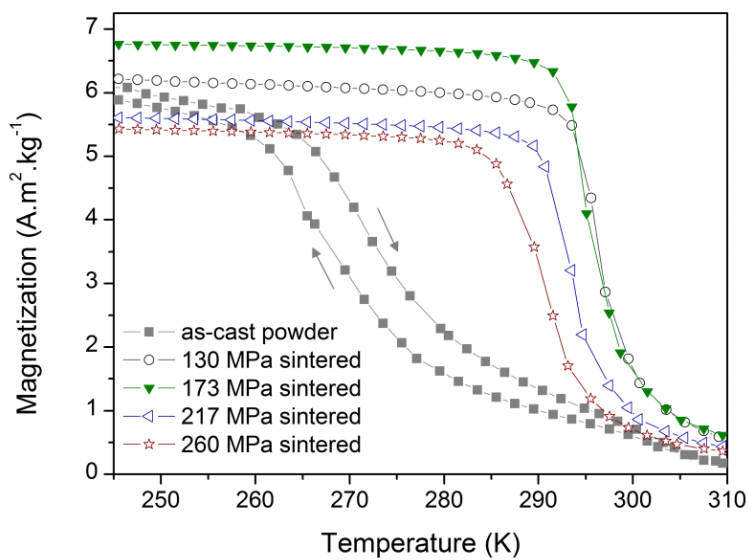

FIG. 2. Magnetization as a function of the temperature for the as-cast powder and the sintered tablets of $Gd_{5.09}Ge_{2.03}Si_{1.88}$ measured at 0.02 T. The sintered tables were measured on cooling.

Analyzing the x-ray diffraction patterns of the as-cast powder, the annealed powder (FIG. 3a) and the sintered tablets (two of them shown in FIG. 3), we have determined the phases and the lattice parameters (Table 1). All the samples present three phases. The monoclinic $Gd_5(Ge,Si)_4$ phase (space group $P112_1/a$), which is responsible for the first-order-magneto-structural transition, is the majority phase (85%) in the as-cast powder. The hexagonal $Gd_5Si_3$ (space group $P63/mcm$) and the orthorhombic GeGd (space group $Cmcm$) phases are minoritary phases. In the annealed powder, the majority phase is the orthorhombic $Gd_5(Ge,Si)_4$ phase (space group $Pnma$), but only 60%. The other phases are the hexagonal $Gd_5Si_3$ (10%) and the orthorhombic GeGd (30%), both present in the as-cast powder. There is no monoclinic $Gd_5(Ge,Si)_4$ phase. The sintered tablets present the same phases as the annealed powder. However, they present significantly more amount of the hexagonal $Gd_5Si_3$ phases and less amount of the orthorhombic $Gd_5(Ge,Si)_4$ phase. The orthorhombic $Gd_5(Ge,Si)_4$ phase is responsible for the second-order magnetic transition observed in the annealed powder (FIG. 1) and in the sintered tablets (FIG. 2). $Gd_5Si_3$, $Gd_5Ge_3$ [14] and $Gd(Si_{0.4}Ge_{0.6})$ [15] compounds have the Néel temperatures at 55, 74, and 57 K, respectively. Then, $Gd_5Si_3$ and GeGd phases are not involved in the second-order transition around room temperature.

Including the annealed powder (zero compacting pressure) in the analyses of the magnetic behavior of the sintered tablets, we have verified that the shift rate of the transition temperature with compacting pressure is -0.025(5) K/MPa, considering a linear behavior (FIG. 4a). It is worth mentioning that no significant variation of the transition temperature with the compacting pressure is observed in the as-pressed (not sintered) tablets. We guess that, in the as-pressed tablets, there is no relevant extra lattice distortions (comparing with the as-cast powder smaller than 50 µm), but there is a residual stress (proportional to the compacting pressure), which is released during the sintering process. This residual stress should be also present in the as-cast powder. Then, annealing the as-cast powder or sintering the as-pressed tablets at 1370 K for 20 hours, this residual stress does not permit the formation of the monoclinic $Gd_5(Ge,Si)_4$ phase and distorts the lattice of the orthorhombic $Gd_5(Ge,Si)_4$ phase, shifting the magnetic transition temperature (higher the compacting pressure, higher the residual stress and lower the transition temperature). While the residual stress in the orthorhombic $Gd_5(Ge,Si)_4$ phase in the tablets leads to a negative shift rate of the transition temperature (-0.025(5) K/MPa), the stress induced by hydrostatic pressure leads to positive shift rates in orthorhombic $Gd_5(Ge,Si)_4$ phase of similar compounds (0.009 K/MPa for $Gd_5Ge_2Si_2$ [16] and 0.003 K/MPa for $Gd_5Ge_{1.9}Si_{2.1}$ [17]).

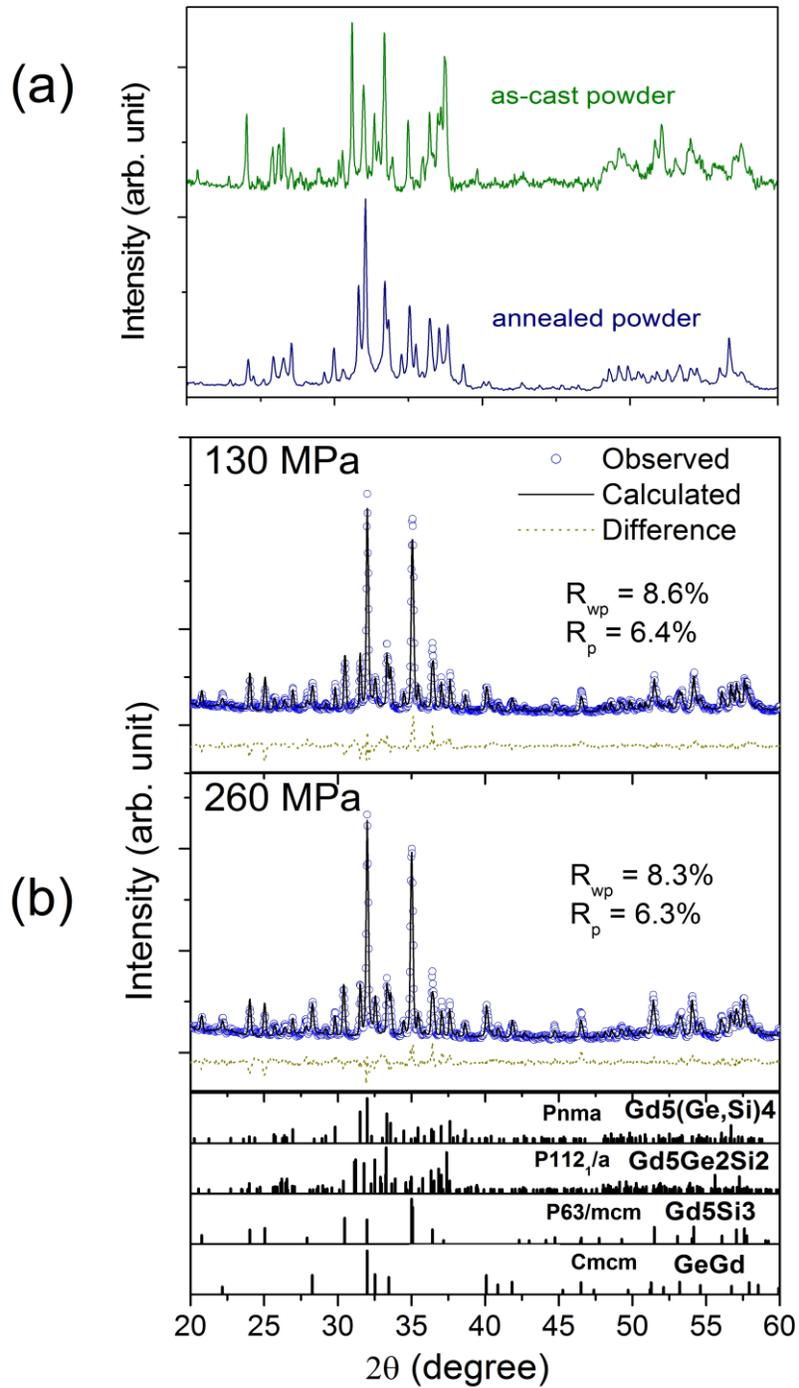

FIG. 3. (a) X-ray diffraction patterns for the as-cast powder and the annealed powder; (b) Rietveld refinement and diffraction pattern for the 130-MPa sintered tablet and 260-MPa sintered tablet and the diffraction peaks for the phases listed in Table 1.

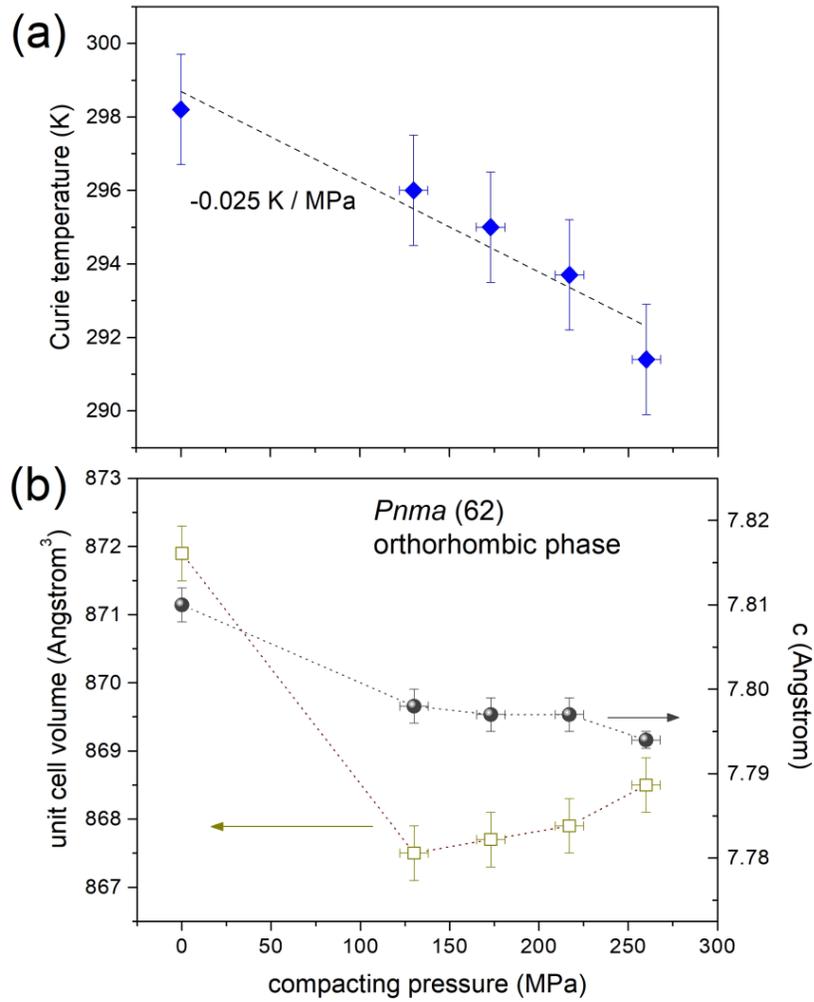

FIG. 4. (a) Curie temperature and (b) unit cell volume and *c* parameter of the *Pnma* orthorhombic phase as functions of the compacting pressure of the powder with subsequent annealing (1370 K / 20 hours).

When we look at the unit cell volume of the orthorhombic $Gd_5(Ge,Si)_4$ phase as a function of the compacting pressure (FIG. 4b), we observe a non-linear behavior. Then, it indicates that there is not a direct relation between magnetic transition temperature and unit cell volume of that phase. Nevertheless, instead of isotropic, there might be an anisotropic correlation between magnetic transition and lattice distortion. The lattice parameters *a* and *b* behave like the unit cell volume as a function of compacting pressure. But, when we look at the behavior of the lattice parameter *c* (FIG. 4b), we observe a tendency not observed for *a, b* nor the unit cell volume. Besides, this tendency resembles the behavior of the Curie temperature ($T_C$) as a function of compacting pressure. This an indicative that $T_C$ is directly affected by parameter $c$ change.

Table1: Structural information obtained from the x-ray diffraction patterns using the Rietveld refinement. The percentages refer to the amount of each phase in the sample; $a$, $b$ and $c$ are the unit cell parameters; $\gamma$ is the angle associated with the monoclinic phase; $V$ is the unit cell volume.

|  | as-cast powder | annealed powder | 130-MPa-sintered tablet | 173-MPa-sintered tablet | 217-MPa-sintered tablet | 260-MPa-sintered tablet |
|---|---|---|---|---|---|---|
| $Gd_5(Ge,Si)_4$ $P112_1/a$ (14:c1) | 85(5)% <br> a = 7.577(1) Å <br> b = 14.817(2) Å <br> c = 7.785(1) Å <br> $\gamma$ = 93.11° <br> V = 872.7(1) Å³ | – | – | – | – | – |
| $Gd_5Si_3$ $P63/mcm$ (193) | 5(2)% <br> a = 8.510(2) Å <br> c = 6.390(1) Å <br> V = 400.8(1) Å³ | 10(2)% <br> a = 8.551(2) Å <br> c = 6.392(1) Å <br> V = 405.0(1) Å³ | 34(2)% <br> a = 8.527(2) Å <br> c = 6.374(1) Å <br> V = 401.4(2) Å³ | 34(2)% <br> a = 8.528(2) Å <br> b = 6.377(2) Å <br> V = 401.6(4) Å³ | 33(2)% <br> a = 8.529(3) Å <br> c = 6.383(2) Å <br> V = 402.1(3) Å³ | 34(2)% <br> a = 8.530(2) Å <br> c = 6.394(1) Å <br> V = 402.9(2) Å³ |
| $Gd_5(Ge,Si)_4$ $Pnma$ (62) | – | 60(2)% <br> a = 7.534(2) Å <br> b = 14.817(3) Å <br> c = 7.810(2) Å <br> V = 871.9(4) Å³ | 42(2)% <br> a = 7.520(2) Å <br> b = 14.792(3) Å <br> c = 7.798(2) Å <br> V = 867.5(4) Å³ | 42(3)% <br> a = 7.522(2) Å <br> b = 14.794(3) Å <br> c = 7.797(2) Å <br> V = 867.7(4) Å³ | 41(3)% <br> a = 7.523(2) Å <br> b = 14.797(4) Å <br> c = 7.797(2) Å <br> V = 867.9(4) Å³ | 35(3)% <br> a = 7.527(1) Å <br> b = 14.803(2) Å <br> c = 7.794(1) Å <br> V = 868.5(4) Å³ |
| GeGd $Cmcm$ (63) | 10(3)% <br> a = 4.186(2) Å <br> b = 10.516(2) Å <br> c = 4.045(1) Å <br> V = 178.1(1) Å³ | 30(3)% <br> a = 4.347(3) Å <br> b = 10.459(7) Å <br> c = 3.998(3) Å <br> V = 181.8(2) Å³ | 24(3)% <br> a = 4.312(1) Å <br> b = 10.686(2) Å <br> c = 3.898(1) Å <br> V = 179.6(2) Å³ | 24(3)% <br> a = 4.311(2) Å <br> b = 10.682(3) Å <br> c = 3.898(2) Å <br> V = 179.5(2) Å³ | 26(3)% <br> a = 4.311(1) Å <br> b = 10.686(2) Å <br> c = 3.898(1) Å <br> V = 179.6(1) Å³ | 31(3)% <br> a = 4.311(1) Å <br> b = 10.687(2) Å <br> c = 3.898(1) Å <br> V = 179.6(1) Å³ |

## 4. Conclusions

We have studied the $Gd_{5.09}Ge_{2.03}Si_{1.88}$ compound processed in different ways, mainly, the as-cast powder, the annealed powder and the pressed and sintered powder. The magnetization curves as a function of the temperature for the as-cast powder and the as-cast bulk have the same profile essentially, in spite of a small $T_C$ shift. This shift is an indicative of a small strain induced by manual grinding. Meanwhile, the annealed powder (1370 K / 20 hours) does not present the monoclinic phase and the first-order-magneto-structural transition observed in the as-cast powder. The pressed and sintered powder also do not present the monoclinic phase and the first-order-magneto-structural transition. Furthermore, the compacting pressure shifts the second-order-magnetic transition to lower temperatures. We guess that a residual stress induced by compacting pressure is released during the sintering process and distorts the lattice of the orthorhombic phase, shifting the second-order-magnetic transition. The behavior of cell parameters as a function of the compacting pressure indicates that Curie temperature is directly affected by parameter *c* change.

**Acknowledgements**

The authors thank financial support from Fundação de Amparo à Pesquisa do Estado de São Paulo – FAPESP and Conselho Nacional de Desenvolvimento Científico e Tecnológico – CNPq.